\documentclass[ip,twocolumn]{jpsj3}
\usepackage{txfonts}
\usepackage{graphicx}

\usepackage{color}
\def\newr{\color{black}}

\def\lsno{La$_{2-x}$Sr$_x$NiO$_4$}
\def\lsco{La$_{2-x}$Sr$_x$CuO$_4$}
\def\lbco{La$_{2-x}$Ba$_x$CuO$_4$}

\def\lnsco{La$_{1.6-x}$Nd$_{0.4}$Sr$_x$CuO$_4$}

\def\ybco{YBa$_2$Cu$_3$O$_{6+x}$}

\def\bscco{Bi$_2$Sr$_2$CaCu$_2$O$_{8+\delta}$}

\title{Superconductivity from charge order in cuprates}

\author{J. M. Tranquada$^1$\thanks{jtran@bnl.gov}, M. P. M. Dean$^1$, and Qiang Li$^{1,2}$}
\inst{$^1$Condensed Matter Physics \&\ Materials Science Division, Brookhaven National Laboratory, Upton, NY 11973-5000, USA \\
$^2$Department of Physics and Astronomy, Stony Brook University, Stony Brook, New York 11794-3800, USA
} 

\abst{Superconductivity in layered cuprates is induced by doping holes into a parent antiferromagnetic insulator.  It is now recognized that another common emergent order involves charge stripes, and our understanding of the relationship between charge stripes and superconductivity has been evolving.  Here we review studies of 214 cuprate families obtained by doping La$_2$CuO$_4$.  Charge-stripe order tends to compete with bulk superconductivity; nevertheless, there is plentiful evidence that it coexists with two-dimensional superconductivity.  This has been interpreted in terms of pair-density-wave superconductivity, and the perspective has shifted from competing to intertwined orders.  In fact, a new picture of superconductivity based on pairing within charge stripes has been proposed, as we discuss.}

\begin{document}
\maketitle

\section{Introduction}

The story of charge-stripe order in cuprate superconductors has evolved in important ways over the last quarter century.  The original discovery of charge stripe order in \lnsco\ \cite{tran95a} was in association with a strong depression of the bulk superconducting transition temperature, and the phase diagrams that have evolved from further investigation suggest competition between stripe order and bulk superconductivity \cite{tran97a}.   The discovery of two-dimensional (2D) superconductivity \cite{li07,tran08} onsetting with spin stripe order, below the charge-stripe transition, has led to an interpretation in terms of pair-density-wave order \cite{berg07,berg09b,agte20} and the concept of intertwined orders \cite{frad15}.  Furthermore, recent experiments have provided evidence that charge stripes are made up of paired holes \cite{li19a} and have motivated the proposal that charge stripes in the form of hole-doped two-leg spin ladders provide the key to understanding pairing in the cuprates \cite{tran21a}.

In this review, we will focus on cuprates in the 214 structural family based on doping the parent correlated insulator La$_2$CuO$_4$.  This structure is shared by a number of transition-metal oxides, and charge-stripe order is common to compounds such as \lsno,\cite{chen93,tran94a,ulbr12b,tran13a} La$_{2-x}$Sr$_x$CoO$_4$,\cite{zali00,ulbr12b} and La$_{1-x}$Sr$_{1+x}$MnO$_4$ \cite{ster96,ulbr12b}.  Among these isostructural compounds, however, superconductivity is unique to cuprates.  We choose to use the name ``charge stripe'' rather than ``charge density wave'' (CDW) to avoid any confusion regarding the origin of the modulation. In principle, both of these names are appropriate to describe a periodic charge modulation in an atomic lattice; however, CDW can carry an association with the Fermi-surface nesting mechanism.  A thoughtful theoretical analysis\cite{joha08} has shown that such a scheme is rarely, if ever, relevant outside of quasi-1D systems.  This applies even more so in cuprates, where the doping of holes into an antiferromagnet causes frustration of both magnetic superexchange between Cu moments and the kinetic energy of the doped holes.  At high temperature, this mixture forms a bad metal; with cooling, the conductivity improves as the holes and spins self-organize into modulated structures.\cite{tran21a}

\begin{figure}[b]
	\centering
	\includegraphics[width=0.5\columnwidth]{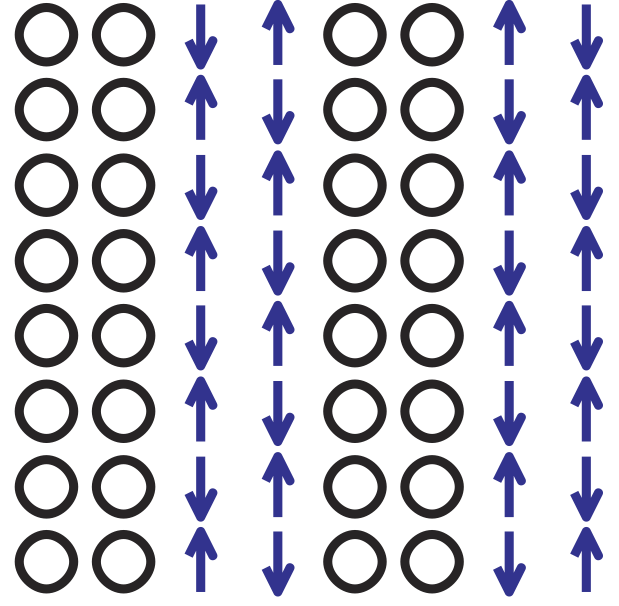}
	\caption{(Color online)  Cartoon of spin and charge stripe orders within a CuO$_2$ plane, where only Cu sites are shown (from Ref.~\citen{tran04}).  The arrows indicate $S=1/2$ moments, which are locally antiferromagnetic, but have an antiphase order across the charge stripes, indicated by circles.  The hole density within a charge stripe is 0.25 per Cu site; there are dynamic spin degrees of freedom on the charge stripes.\cite{tran21a}}
	\label{fg:stripes}
\end{figure}

A cartoon of charge stripe order at a doping of $x=1/8$ is presented in Fig.~\ref{fg:stripes}; for simplicity, only Cu sites within a CuO$_2$ plane are represented.  The arrows indicate the antiferromagnetic (AF) arrangement of Cu moments; however, note that the AF phase shifts by $\pi$ on translating from one spin stripe to the next.  This antiphase relationship is important for decoupling from the charge and spin degrees of freedom on a charge stripe.\cite{tran21a}  Within this context, the charge stripes can be viewed as doped 2-leg spin-1/2 ladders, where the spins have singlet correlations and the hole density is $\sim0.25$.

We will begin by considering the lattice symmetry and its influence on the observation of charge-stripe order.  We will then discuss evidence for dynamic stripes and their coupling to phonons.  Finally, we will describe the evidence that stripes coexist with 2D superconductivity and that the holes in charge stripes tend to respond as pairs.  The implications of these results will be explored.

\section{Charge-stripe order and lattice symmetry}

The first clue came with the measurement of the superconducting transition temperature, $T_c$, as a function of doping in \lbco, where a strong dip was found\cite{mood88} at $x\approx1/8$.  Soon after, x-ray diffraction measurements demonstrated a low-temperature structural phase transition\cite{axe89,axe89b} not present in \lsco.  At high temperature, these compounds have the K$_2$NiF$_4$ structure (space group $I4/mmm$), with a centered stacking of CuO$_2$ layers, which are formed from corner-sharing CuO$_6$ octahedra (with Jahn-Teller distortion); this is commonly referred to as the high-temperature tetragonal (HTT) structure.  On cooling, there is a transition to the low-temperature orthorhombic (LTO) phase (space group $Bmab$), in which the octahedra rotate about a diagonal axis such that all nearest-neighbor Cu-O bonds remain equivalent.  The new transition in LBCO is from the LTO phase to a low-temperature tetragonal (LTT) phase (space group $P4_2/ncm$), in which the tilt direction of the octahedra rotates to be along a Cu-O bond direction.  The Cu-Cu spacing is the same in orthogonal directions (tetragonal symmetry), and the Cu-O bonds along the rotation axis are parallel to the plane, which means that the Cu-O bonds along the tilt direction are expanded.  This anisotropy between orthogonal Cu-O bonds rotates by 90$^\circ$ from one layer to the next.

Next, it was discovered that partial substitution of a rare-earth ion, such as Nd$^{3+}$, that is smaller than the La$^{3+}$, can induce in LSCO the LTT phase\cite{craw91}; it can also induce an intermediate phase, the low-temperature less-orthorhombic (LTLO, space group $Pccn$) phase, with a mixture of the two types of octahedral tilts.  A study of La$_{2-x-y}$Nd$_y$Sr$_x$CuO$_4$ demonstrated that the occurrence of the LTT phase with sufficiently large octahedral tilts correlated with a strong suppression of bulk superconductivity.\cite{buch93,buch94a}  Such observations motivated the growth of single crystals of La$_{1.6-x}$Nd$_{0.4}$Sr$_x$CuO$_4$ (LNSCO) and the investigation of their anisotropic transport properties.\cite{naka92}  

Several Hartree-Fock analyses of the two-dimensional (2D) Hubbard model with parameters relevant to hole-doped CuO$_2$ planes found inhomogeneous solutions involving charge and spin stripes \cite{zaan89,schu89,mach89}.  In each case, the hole-density within the charge stripes was one hole per Cu site, corresponding to insulating behavior, but the antiferromagnetic spin order was incommensurate, with a period twice that of the charge order.  There were also predictions of inhomogeneous phases\cite{low94} based on models of frustrated phase separation in cuprates.\cite{emer93,kive94}   Low-energy incommensurate spin excitations were discovered in superconducting LSCO,\cite{cheo91} with an orientation consistent with spin stripes and a wave vector that grew with doping, but a period that would require putative charge stripes to have a hole density of just a half per Cu site.  Following the discovery of combined charge and spin stripe orders in La$_2$NiO$_{4.125}$,\cite{tran94a} and with knowledge of the dip in $T_c$ at 1/8 doping in LBCO\cite{mood88} and LNSCO,\cite{buch94a} there was motivation to take advantage of the single crystals of LNSCO\cite{naka92} to look for possible stripe order.  Indeed, neutron diffraction on a crystal of LNSCO with $x=0.12$ provided evidence for spin stripe order with incommensurate peaks split about the antiferromagnetic wave vector, ${\bf Q}_{\rm AF}=(0.5,0.5,0)$, by ${\bf q}_{\rm so} = (\epsilon,0,0)$ in reciprocal lattice units (for the HTT phase), with $\epsilon=0.12\approx x$, and charge order with ${\bf q}_{\rm co} = (2\epsilon,0,0.5)$.\cite{tran95a}  The charge-order peak intensity appeared just below the structural transition from the HTT phase toward the LTT phase, with the spin order developing at a slightly lower temperature.

\begin{figure}[t]
	\centering
	\includegraphics[width=0.8\columnwidth]{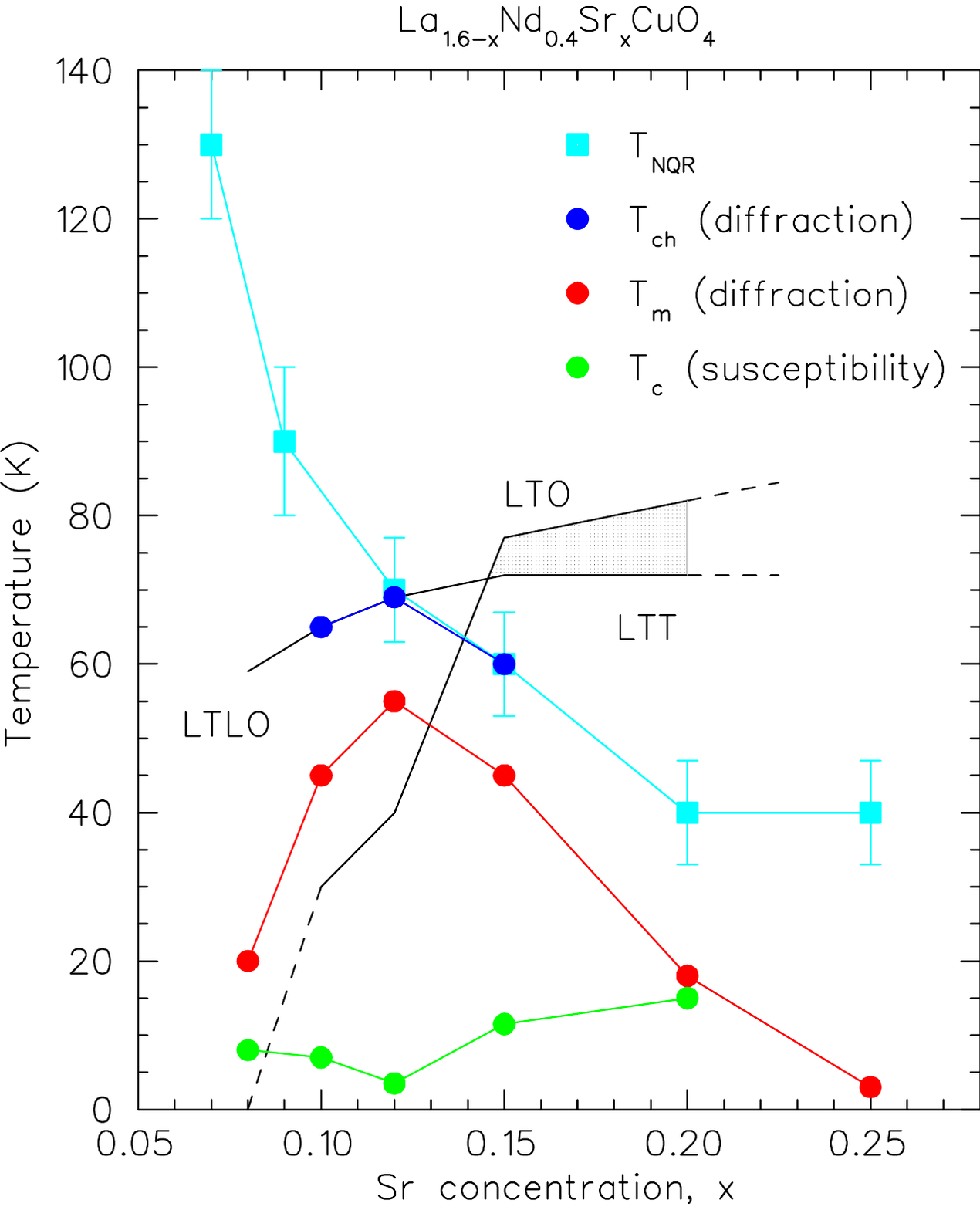}
	\caption{(Color online)  Temperature vs.\ hole-doping phase diagram for \lnsco\ from Ref.~\citen{ichi00}.  $T_{\rm NQR}$ (cyan squares) indicates onset of wipeout effect.\cite{sing99}  Neutron results for onset of charge order (spin order) are indicated by dark blue (red) circles.  Green circles denote $T_c$ from magnetic susceptibility.  Structural transitions between HTT, LTLO, and LTT are indicated by black lines.  }
	\label{fg:ndpd}
\end{figure}

\begin{figure}[t]
	\centering
	\includegraphics[width=0.9\columnwidth]{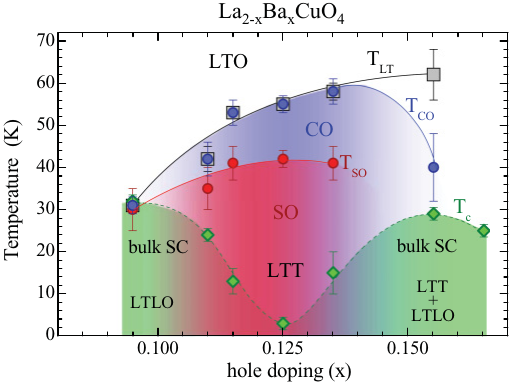}
	\caption{(Color online)  Temperature vs.\ hole-doping phase diagram for \lbco\ from Ref.~\citen{huck11}.  Symbols indicate transition temperatures for structure (grey squares), charge-stripe order (blue circles), spin-stripe order (red circles), and superconductivity (green diamonds).}
	\label{fg:bapd}
\end{figure}

A series of single-crystal neutron diffraction studies\cite{tran95a,tran96b,tran97a,tran99a,ichi00,waki03} on LNSCO eventually led to the phase diagram shown in Fig.~\ref{fg:ndpd}; confirmation of these results and completion at large $x$ have been provided by recent work.\cite{drag20,ma21}  Work on LBCO took a bit longer to realize, because of the challenge in growing crystals.  Once crystals with $x=1/8$ were successfully grown, neutron diffraction\cite{fuji04} demonstrated charge and spin stripe order similar to that in LNSCO.  A combination of neutron and hard x-ray diffraction measurements\cite{kim08a,huck11} eventually led to the phase diagram shown in Fig.~\ref{fg:bapd}.  Hard x-rays have also been used to study the evolution of charge order at $x=1/8$ with hydrostatic pressure, where the charge order was found to survive the transition from LTT to HTT.\cite{huck10}  X-rays were also used to follow the charge order in a $c$-axis magnetic field; away from $x=1/8$, where the charge was not saturated in zero field, the charge order increased with applied field,\cite{huck13} while at $x=1/8$ fields above 5~T applied at $T=2$~K cause a gradual increase in correlation length.\cite{kim08b} 

Neutron diffraction detects the charge modulation indirectly, through the modulated atomic displacements inevitably associated with it; the situation is similar for diffraction with hard x-rays.  In contrast, resonant soft--x-ray scattering (RSXS) performed at the O $K$ edge or Cu $L_3$ edge couples directly to the electronic states whose occupancy is modulated.  As first demonstrated for LBCO $x=1/8$,\cite{abba05} the charge order peak is detected only when the photon energy is tuned to the pre-edge peak of the O $K$ edge, corresponding to electronic excitations into empty O $2p$ states,\cite{chen91} or to the leading edge of the large peak at the Cu $L_3$ edge, corresponding to transitions into the Cu $3d_{x^2-y^2}$ hole.\cite{bian88}  Such measurements have been reproduced in LBCO\cite{wilk11,tham13} and LNSCO.\cite{wilk11,achk13}  A study of the polarization dependence of the O $K$ edge and Cu $L_3$ edge charge-order scattering from LBCO indicates that the structure factor has $s'$ symmetry,\cite{achk16b} which is consistent with a sinusoidal modulation of the hole density in O $2p$ orbitals.

We have already noted that the anisotropy that pins the stripes within the CuO$_2$ planes rotates 90$^\circ$ from one layer to the next.  This respects the crystal symmetry, such that the intensity of $(00L)$ reflections is finite only for even $L$.  In RSXS experiments, however, the x-ray polarization can break the symmetry,\cite{temp80} resulting in finite intensity at (001) when measured at a suitable absorption resonance.\cite{achk16a}   Measurements on 1/8-doped LBCO and LNSCO show that the (001) intensity turns on rather sharply below the transition to the LTT phase, where the charge-order superlattice peaks also appear.  Results for La$_{1.8-x}$Eu$_{0.2}$Sr$_x$CuO$_4$ with $x=0.15$ also show the onset of (001) at the LTT transition; however, in this case the charge stripe diffraction appears at a considerably lower temperature.\cite{achk16a}  The broken rotational symmetry without actual charge-stripe order is evidence of nematic order, a state of oriented but fluctuating stripes predicted to occur in cuprates.\cite{kive98}  Such measurements have recently been applied to following the disappearance of charge order in LNSCO at large $x$.\cite{gupt21}

When spin-stripe order occurs, it is generally observed to appear at $T<T_{\rm co}$.  As a consequence, techniques that are sensitive to spin order or dynamics may also provide information relevant to charge order.  For example, muon spin resonance ($\mu$SR) in zero applied field can detect a local hyperfine field due to electronic spin order, and $\mu$SR studies provided early confirmations of the stripe order in LBCO and LNSCO.\cite{nach98}  Given that the technique is sensitive even for polycrystalline samples, the first evidence for stripe order in LESCO came from $\mu$SR,\cite{klau00} where spin order was observed for $x$ out to 0.2, with the maximum ordering temperature at $x=0.12$.  Charge-stripe order was later confirmed by RSXS on single crystals.\cite{fink09,fink11,zwie16}

Stripe order in LESCO was expected because it has a large LTT regime.\cite{buch94b}  In contrast, LSCO was long assumed to have the LTO phase down to low temperature.\cite{rada94}  While dynamic incommensurate spin correlations occur over a broad doping range,\cite{cheo91,yama98} it was more of a surprise when charge-stripe order was first inferred from nuclear magnetic resonance (NMR) measurements\cite{hunt99,juli99} and eventually confirmed for $x\sim0.12$ by hard x-ray diffraction.\cite{crof14,tham14}  It has been pointed out that NMR is directly sensitive to magnetic correlations\cite{juli03}; nevertheless, neutron scattering on LBCO $x=1/8$ has demonstrated that the low-energy magnetic fluctuations become quite strong as soon as the charge order sets in,\cite{tran08} and a reinvestigation of LSCO $x=0.115$ by NMR has confirmed that the onset of the ``wipeout'' of the signal associated with $^{63}$Cu nuclear quadrupole resonance is associated with the onset of charge order.\cite{imai17,imai18}

\begin{figure}[t]
	\centering
	\includegraphics[width=0.9\columnwidth]{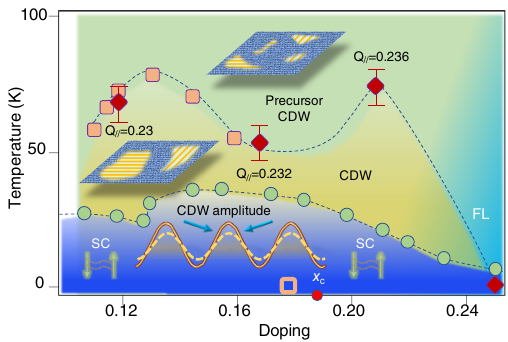}
	\caption{(Color online)  Temperature vs.\ hole-doping phase diagram for \lsco, from Ref.~\citen{miao21}.  Red diamonds and pink squares indicate temperatures below which the charge-stripe correlation length begins to grow.  Green circles denote $T_c$.  }
	\label{fg:srpd}
\end{figure}

The pinning of charge stripes has now been explained by the neutron diffraction observation of weak diffraction peaks in a single crystal of LSCO $x=0.07$ that indicate the structure to be LTLO rather than LTO.\cite{jaco15}  The deviation from LTO is small, but appears to survive up to the transition to the HTT phase.  RSXS studies\cite{wu12,wen19,miao21} on LSCO have demonstrated that the onset of charge stripe correlations extends out to at least $x=0.21$, as shown in Fig.~\ref{fg:srpd}. (The $x=0.21$ crystal studied in Ref.~\citen{miao21} has been shown to be orthorhombic below 240~K by unpublished single-crystal neutron diffraction measurements.)  There is an interesting change in temperature dependence with doping, however.  For $x\sim0.12$, the charge order peak intensity starts growing below $\sim70$~K and continues to grow on cooling below $T_c$; in contrast, the intensity drops below $T_c$ for $x>0.14$.\cite{wu12,wen19,miao21}   This is evidence of the competition between stripe order and spatially-uniform superconductivity, which will be discussed below.

\section{Stripe dynamics and coupling to phonons}

Spin-stripe fluctuations are readily measured by inelastic neutron scattering with an energy resolution of $\sim 1$~meV, and have been characterized in a number of 214 cuprates, as reviewed elsewhere.\cite{birg06,fuji12a}  In principle, one should also be able to measure the lattice fluctuations associated with charge stripes, but the signal strength has proven to be a challenge.  Nevertheless, the possibility has been demonstrated in \lsno.\cite{anis14,zhon17}  One can also get good energy resolution with inelastic x-ray scattering using hard x-rays.  A study of LBCO $x=1/8$ found anomalous features in low-energy phonons crossing ${\bf q}_{\rm co}$ at temperatures up to 300~K.\cite{miao18}

\begin{figure}[t]
	\centering
	\includegraphics[width=0.8\columnwidth]{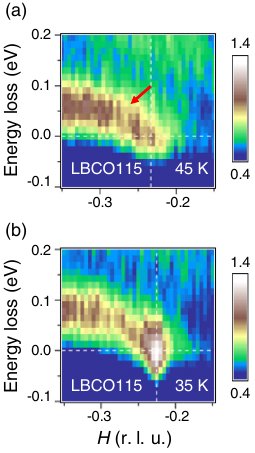}
	\caption{(Color online)  Cu $L_3$ edge RIXS intensity maps of excitations vs.\ momentum transfer along $(H,0,0)$ for LBCO $x=0.115$ at (a) 45 K, (b) 35 K.  From Ref.~\citen{miao19}.}
	\label{fg:disp}
\end{figure}

A successful alternative technique is resonant inelastic x-ray scattering (RIXS) using soft x-rays.  This is a rapidly developing technique, where the energy resolution was a challenge but is rapidly improving.  An initial study of LBCO $x=1/8$ with an energy resolution of 90~meV demonstrated that dynamic charge stripes are present at temperatures as high as 90~K.\cite{miao17}  The incommensurability $2\epsilon$ obtained for the fluctuating charge stripes in the LTO phase becomes larger, in contrast to the decrease in $\epsilon$ for the spin fluctuations measured with neutrons.\cite{fuji04}  Similar RIXS results for LESCO $x=1/8$ have been reported.\cite{wang20a}

A new experiment on LBCO $x=0.115$ using an energy resolution of 55~meV is much more revealing,\cite{miao19} as indicated in Fig.~\ref{fg:disp}.  At 35~K, below $T_{\rm co}$, one can see that there is strong elastic weight together with excitations dispersing up to $\sim70$~meV at larger $Q$.  Related measurements on LESCO have also appeared.\cite{peng20,wang21}  The excitations are presumably due largely to phonons weighted (and softened) by the coupling to the modulated Cu $3d$ holes.  Indeed, neutron scattering measurements of the optical bond-stretching phonon revealed a strong softening and broadening at ${\bf q}_{\rm co}$, where the phonon energy is $\sim85$~meV at zone center and $\sim70$~meV at zone boundary.\cite{rezn06}  The neutron studies show similar results in LSCO and LNSCO, and it has been reproduced in LBCO $x=0.14$ with non-resonant inelastic x-ray scattering.\cite{dast08}  Using a combination of neutron and x-ray scattering, the doping dependence of the softening at ${\bf q}_{\rm co}$ was studied in LSCO,\cite{park14} and it was found to be present throughout the doping range where superconductivity is strong.

\begin{figure}[t]
	\centering
	\includegraphics[width=0.8\columnwidth]{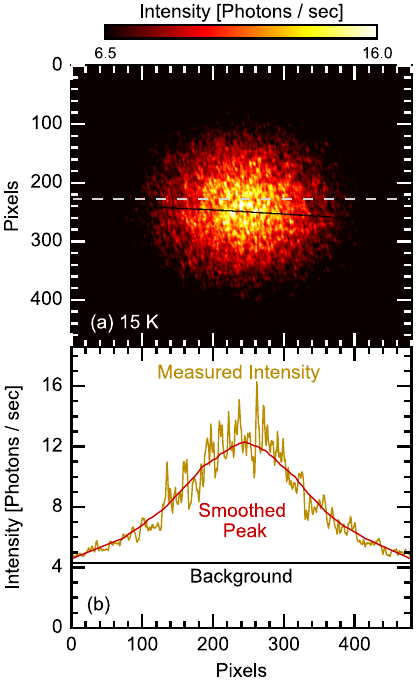}
	\caption{(Color online)  (a) Detector image for diffraction at ${\bf q}_{\rm co}$ from LBCO $x=1/8$ at 15 K.  The black diagonal line is an artifact from the beamstop. (b) Line cut of the measured intensity (oscillating yellow line) along the dashed line in (a). Red and black lines indicate the smoothed peak envelope and the fluorescence-dominated background, respectively. 
	From Ref.~\citen{chen16}.}
	\label{fg:speck}
\end{figure}

It is clear that energy resolution may impact the apparent character of dynamic correlations.  There are certainly fluctuations over a substantial energy scale, as shown in Fig.~\ref{fg:disp}.  Given that every scattering technique has a finite energy resolution, one might ask whether charge stripes are ever truly static and ordered.  To answer this question, we turn to x-ray photon correlation spectroscopy (XPCS).  This technique is only possible with the use of a highly coherent photon source.\cite{sutt08}  Measuring diffraction from a sample with a domain structure using coherent x-rays, an image of the diffracted beam has the form of a complex speckle pattern.  Any changes in order with time will cause a variation in the speckle interference pattern.  By monitoring the speckle pattern as a function of time, one can test for fluctuations in the underlying order.

Figure~\ref{fg:speck}(a) shows the speckle pattern due to diffraction from charge stripes in LBCO $x=1/8$.\cite{chen16}  If one takes a cut through the pattern, as in (b), and plots it vs.\ time, changes in the interference due to fluctuations will become apparent.  For LBCO $x=0.125$ and 0.11, no significant changes were observed over a time scale of $\sim2$~h.\cite{chen16,tham17}  This is strong evidence for the static character of charge-stripe order.

\begin{figure}[t]
	\centering
	\includegraphics[width=0.8\columnwidth]{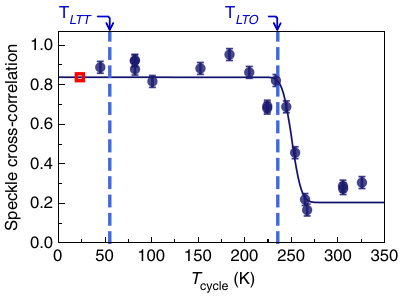}
	\caption{(Color online)  Measurements of the cross correlation for speckle patterns measured at ${\bf q}_{\rm co}$ in LBCO $x=1/8$ before and after cycling the temperature to $T_{\rm cycle}$.  The vertical dashed lines indicate the temperatures of the LTT-LTO and LTO-HTT transitions. From Ref.~\citen{chen19b}.}
	\label{fg:mem}
\end{figure}

XPCS studies can also provide information on factors that control the pinning of charge stripes.  For example, suppose one measures the diffraction speckle pattern at low temperature, then heats the sample to $T_{\rm cycle}$, and finally cools to the original temperature and repeats the speckle measurement.  By evaluating the cross correlation of the speckle patterns, one can test the reproducibility of the stripe domain pattern.  Figure~\ref{fg:mem} shows the results of such a study on LBCO $x=1/8$.\cite{chen19b}  The speckle correlations are insensitive to cycling into the LTO phase but are drastically affected by cycling into the HTT phase.  The local orientation of stripes is sensitive to the LTT twin domain pattern, so the results suggest that the LTT pattern reproduces quite well from cycling into the LTO phase, consistent with transmission electron microscopy work.\cite{zhu94,chen93}  This situation is disrupted by cycling into the HTT phase, which presumably leads to changes in the LTO (and hence LTT) twin-domain pattern.

\section{Evidence for superconducting charge stripes}
\label{SCstripes}

\begin{figure}[t]
	\centering
	\includegraphics[width=0.8\columnwidth]{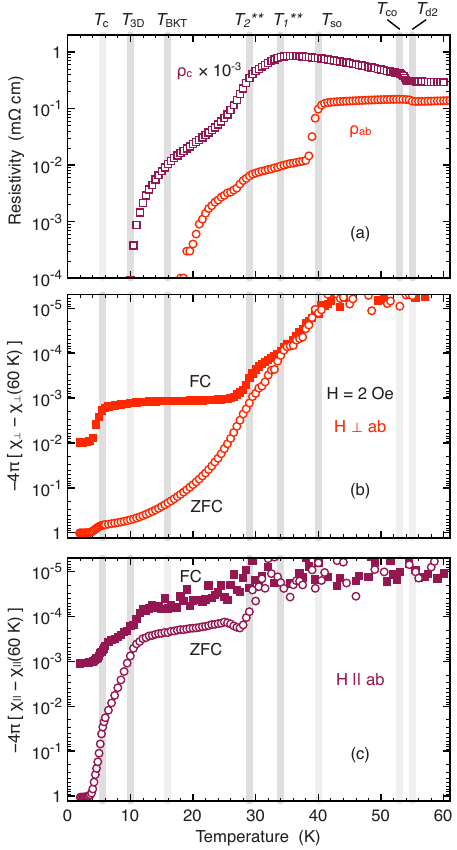}
	\caption{(Color online)  Temperature-dependent measurements on single crystals of LBCO $x=1/8$.  (a) In-plane, $\rho_{ab}$ (circles), and $c$-axis, $\rho_c$ (squares), resistivities.  (b) Magnetic susceptibility measured in field of 2 Oe applied perpendicular to the planes for field-cooled (FC, filled squared) and zero-field-cooled (ZFC, open circles) conditions.  (c) Magnetic susceptibility for field parallel to the planes.  From Ref.~\citen{tran08}.}
	\label{fg:2dsc}
\end{figure}

The phase diagrams in Figs.~\ref{fg:ndpd} and \ref{fg:bapd} clearly show that charge stripe order competes with 3D superconductivity; however, it does not follow that stripes are bad for pairing.  In fact, careful measurements of anisotropic resistivity, shown in Fig.~\ref{fg:2dsc}(a), find that there is an order of magnitude decrease in the in-plane resistivity, $\rho_{ab}$, at 40~K, coincident with the spin-stripe ordering and below $T_{\rm co}$.\cite{li07,tran08}  The resistivity perpendicular to the planes, $\rho_c$, shows no change at 40~K, and it is still quite significant at 16~K, where nonlinear $I$-$V$ data are consistent with a Kosterlitz-Thouless transition, indicating 2D superconducting phase order.  The possibility that the drop in $\rho_{ab}$ at 40~K is the onset of phase-disordered 2D superconductivity is supported by measurements of the magnetic susceptibility, which show an onset of weak diamagnetism at that temperature when the response is from the planes, measured with $H\perp ab$ as in Fig.~\ref{fg:2dsc}(b), but not when interlayer Josephson currents are required, as in Fig.~\ref{fg:2dsc}(c).

To explain the observation of 2D superconductivity, it has been proposed that the charge stripes are superconducting, but coupled in an antiphase fashion, corresponding to a pair-density-wave (PDW) state.\cite{berg07,berg09b,agte20}  The modulation of the amplitude of the pair wave function means that it passes through zero at the center of each spin stripe, consistent with the empirical observation that superconductivity does not locally coexist with AF order in cuprates.  Because the orientation of the PDW should follow the stripes and rotate by $90^\circ$ from one layer to the next, the interlayer Josephson coupling is frustrated, thus inhibiting 3D superconducting order.  While there have been proposals that fluctuating charge stripes might be the key to high-temperature superconductivity,\cite{cast95,capr17}  the observation of 2D superconductivity has shifted the conversation from one of competing orders to one of intertwined orders.\cite{frad15}

{\newr
The appearance of 3D superconducting order below $\sim5$~K has a filamentary character.  To understand it, we need to take account of the fact that the charge-stripe order has a finite correlation length.  It is likely determined by disorder in the local hole density due to poor screening of the the long-range Coulomb interaction associated with disordered dopants.\cite{tran21a}   In particular, NMR studies of LSCO provided evidence for a hole-density distribution width of 0.05 at $x=0.15$.\cite{sing02a,sing05}  Hence, we expect to have small patches of higher hole concentration, where uniform $d$-wave order can develop.  Coupling of such rare domains in neighboring layers along the $c$ axis will only develop percolating coherence at low temperature.  Also, we expect the uniform $d$-wave regions to compete with the PDW order (which coexists with spin-stripe order), so the filamentary 3D order should be independent of the 2D SC.
}

Complementary support for the PDW picture is provided by optical conductivity studies.  For example, measurements of the loss of the $c$-axis superconducting response in La$_{1.85-y}$Nd$_y$Sr$_{0.15}$CuO$_4$ as the structure was tuned from LTO to LTT by increasing the Nd concentration\cite{taji01} led to an initial proposal for the PDW state.\cite{hime02}  Optical conductivity results for \lbco\ $x=1/8$ reveal a partial shift of conductivity from the Drude peak to energies above 40~meV on cooling below the onset of 2D superconductivity, which is consistent with predictions for a modulated superconductor.\cite{home06,home12}  Angle-resolved photoemission measurements on superconducting LBCO samples reveal a spectral function that resembles that of other cuprate superconductors, except for the absence of a sharp quasiparticle peak even at the nodal wave vector.\cite{vall06,vall07,he09}  This is likely due to the presence of the spin-stripe order as discussed in Ref.~\citen{jaco15}.

\begin{figure}[t]
	\centering
	\includegraphics[width=0.8\columnwidth]{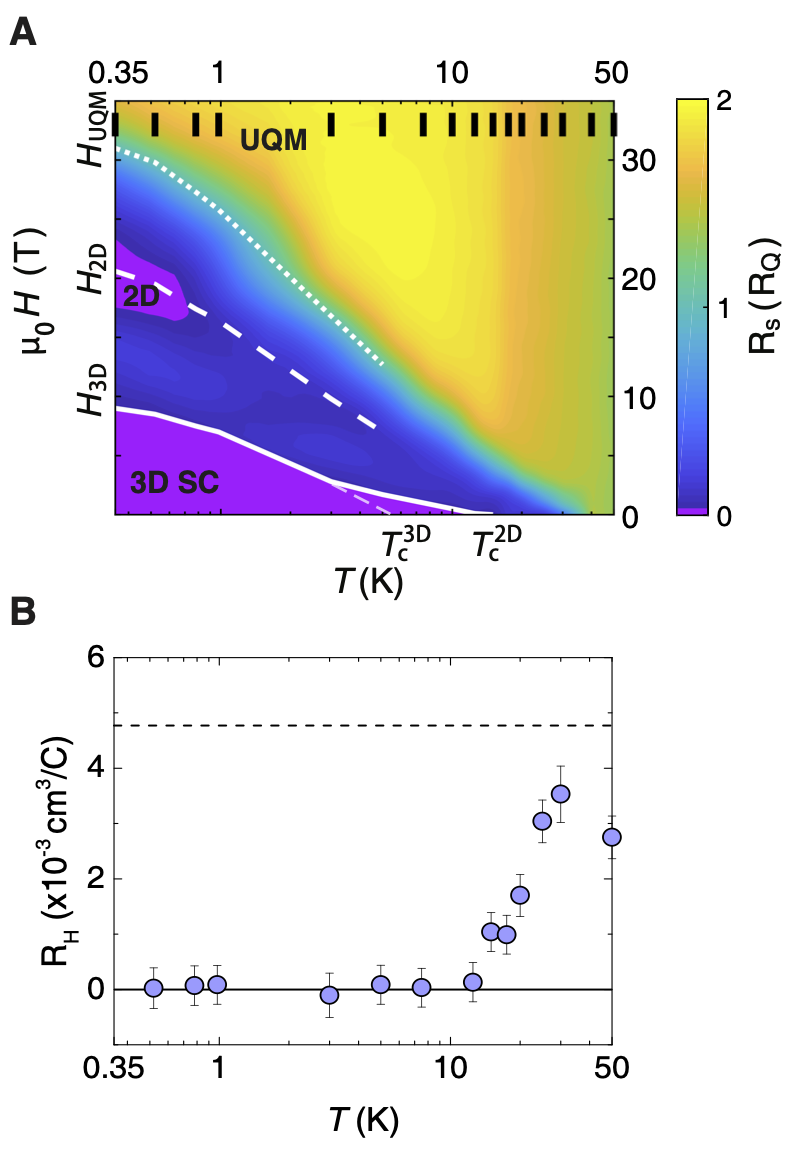}
	\caption{(Color online)   (A) Magnetic field vs.\ temperature phase diagram of LBCO $x=1/8$ in terms of sheet resistance, $\rho_{ab}/(0.5c)$ (where $0.5c$ is the layer separation), in units of the quantum of resistance for pairs, $R_{\rm Q} = \hbar/(2e)^2$; note that the numeric labels are at the top and on the right, and the $T$ scale is logarithmic.   The regimes of 3D and 2D superconductivity (SC) with zero electrical resistance are labeled; the UQM phase occurs at fields above the dotted line. Characteristic fields $H_{\rm 3D}$, $H_{\rm 2D}$, and $H_{\rm UQM}$ (defined in the text) are overplotted as solid, dashed, and dotted white lines, respectively. (B) Hall coefficient as a function of temperature, with error bars obtained by averaging over the entire field range (0 to 35 T).  $R_{\rm H}$ is effectively zero below 15 K, as expected for a superconductor, and it rises to the normal-state magnitude around $\sim40$~K. The upper dashed line indicates the magnitude of $R_{\rm H}$ that would be expected in a one-band system with a nominal hole density of 0.125.  From Ref.~\citen{li19a}.}
	\label{fg:hf}
\end{figure}

Further evidence for pairs in charge stripes comes from transport measurements in high magnetic fields.  Figure~\ref{fg:hf}(A) shows the in-plane resistance as a function of temperature (on a log scale) and magnetic field (up to 35~T).\cite{li19a}  Consider the behavior at base temperature.  The 3D superconductivity ends when finite resistivity appears at $H_{\rm 3D} \approx 10$~T, but the resistivity eventually drops to zero again in a regime of reentrant 2D superconductivity centered on $H_{\rm 2D}\approx 20$~T. (Related results have been reported for LNSCO and LESCO.\cite{shi20a,shi20b})  That order is lost at higher field, but where one might expect the resistivity to diverge toward insulating character, it actually saturates at a surprising value, corresponding to a sheet resistance equal to twice the quantum of resistance for pairs.  This unusual behavior has been labelled the ultra-quantum metal (UQM) state.

Complementary information on the Hall coefficient, $R_{\rm H}$, is present in Fig.~\ref{fg:hf}(B).  At each temperature, $R_{\rm H}$ is averaged over the entire field range; variations with field are within the error bars.  As one can see, $R_{\rm H}\approx0$ in the superconducting state, as expected for the particle-hole symmetry of pairs.  The surprising thing is that this symmetry is still present in the UQM phase.  One possibility is that, even after the loss of superconducting phase coherence between charge stripes at high field, the holes in the stripes still behave as pairs.  The UQM phase is potentially a Bose metal state.\cite{ren20}

A mechanism for superconducting stripes has been proposed.\cite{tran21a}  The charge stripes act as two-leg spin-1/2 ladders doped with holes.  Theorists have demonstrated that such doped ladders should result in pairing.\cite{dago92,dago96}  One challenge is to isolate the spin and charge degrees of freedom from the surrounding environment.  Lessons\cite{boot03b,merr19} from \lsno\ show that antiphase spin-stripe order decouples them from the charge stripes.\cite{tran21a}  To achieve superconducting order, it is necessary to have Josephson coupling between neighboring charge stripes.\cite{emer97}  Because the pairs do not readily coexist with local AF order, antiphase coupling is the favored option, resulting in PDW order.  

The stripe correlations can still be relevant to pairing in the case of spatially-uniform superconductivity.  To obtain in-phase superconductivity between neighboring stripes, it is necessary to gap the spin excitations within the spin stripes.  This implies that the superconducting gap associated with long-range coherent superconducting order is limited by the spin gap, a correlation that has been experimentally demonstrated.\cite{li18}  There is also empirical evidence that the spin gap on the charge stripes provides an upper limit for the pairing scale.\cite{tran21a}

Theory has not yet provided a conclusive answer on this story.  Advanced numerical simulation techniques have provided evidence for pairing within charge stripes,\cite{whit09,jian20} although achieving long-range superconducting correlations is a challenge.\cite{whit15}  PDW order is among the states that are energetically competitive,\cite{hime02,racz07,corb14} but the lowest-energy state can be sensitive to the choice of Hamiltonian.\cite{zhen17}    Perhaps this should not be surprising, since realization of PDW order in cuprates requires a lattice anisotropy capable of pinning the stripe orientation, and uniform $d$-wave superconductivity is more common.

\section{Further connections}

\begin{figure}[t]
	\centering
	\includegraphics[width=0.8\columnwidth]{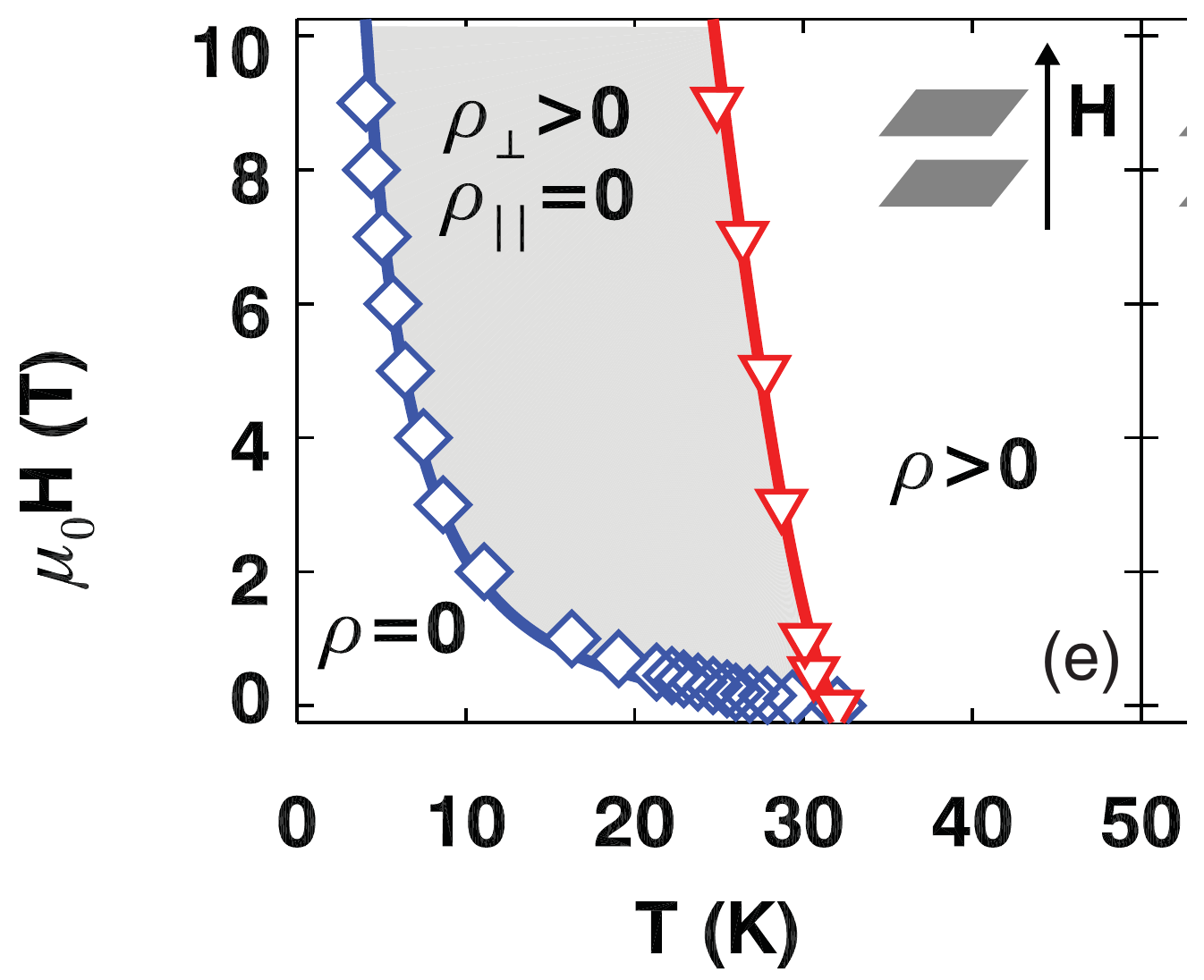}
	\caption{(Color online)  Magnetic field vs.\ temperature phase diagram for LBCO $x=0.095$ for field perpendicular to the planes.  Diamonds indicate the onset of finite $c$-axis resistivity ($\rho_\perp >0$), triangles denote in-plane resistivity becoming non-zero ($\rho_\| > 0$).  The shaded region corresponds to 2D superconductivity.  From Ref.~\citen{wen12b}.}
	\label{fg:magf}
\end{figure}

While $T_c$ shows a maximum of 32~K in LBCO at $x=0.095$, applying a $c$-axis magnetic field can easily destroy the 3D order, resulting in 2D superconductivity,\cite{wen12b} as shown in Fig.~\ref{fg:magf}.  Given the presence of stripe order\cite{huck11} and its enhancement by a magnetic field,\cite{huck13} it is reasonable to assume that, as the magnetic field suppresses the 3D order associated with spatially-uniform superconductivity, PDW order takes over.  This phase diagram has been extended to fields up to 35~T, where 2D superconductivity is still present below 12 K.\cite{steg13}

\begin{figure}[t]
	\centering
	\includegraphics[width=0.8\columnwidth]{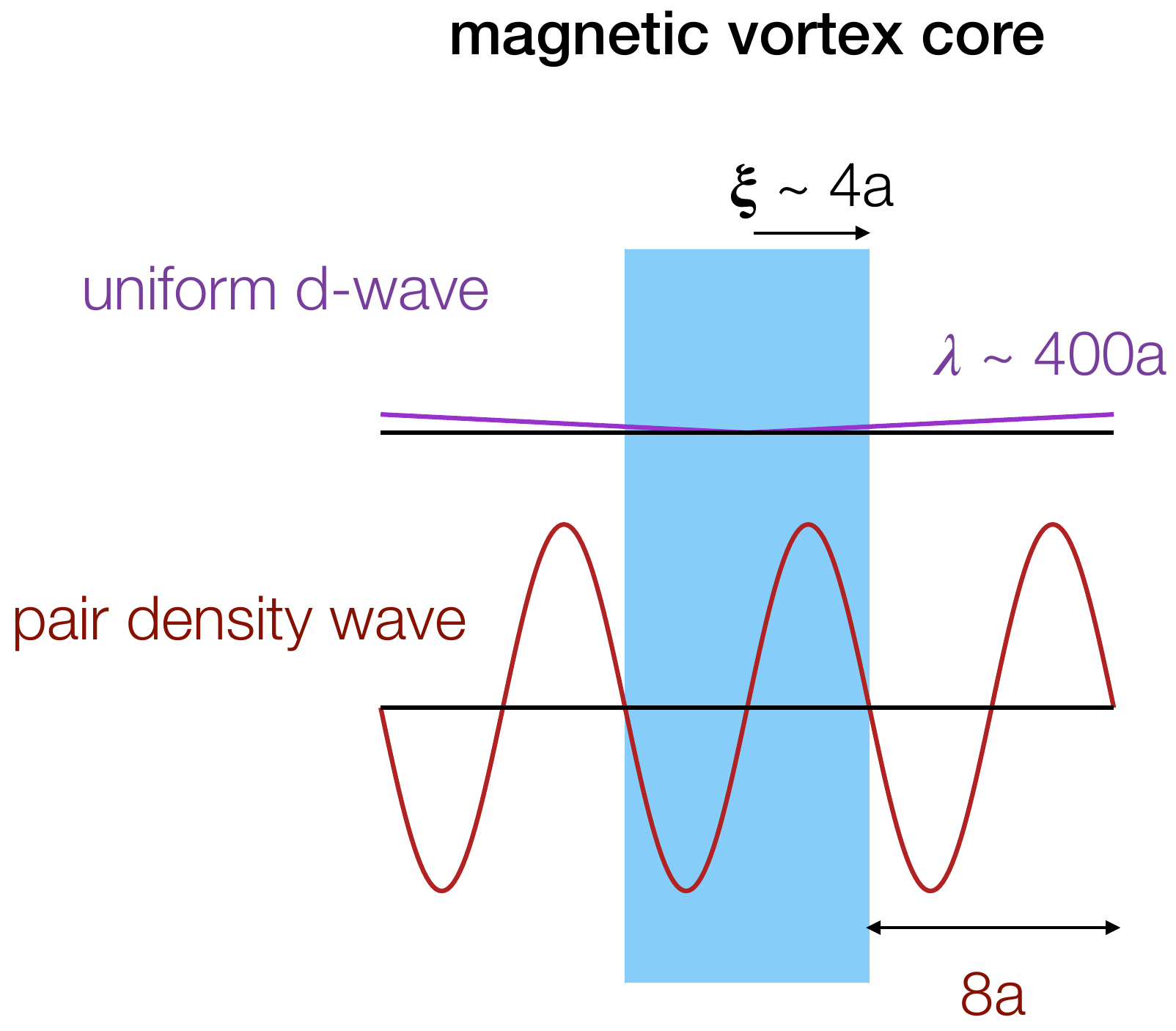}
	\caption{(Color online)  A magnetic vortex core has a radius equal to the superconducting coherence length, which for cuprates is of order 4$a$, where $a$ is the Cu-Cu lattice spacing.  The superconducting order is suppressed within the core, and outside the core it recovers on the scale of the magnetic penetration depth $\lambda\sim400a$.  In contrast, the PDW oscillates with a period of $8a$, which may provide an energetic advantage near a vortex core, as discussed in the text.}
	\label{fg:vc}
\end{figure}

A connection between magnetic vortices and charge order has also been seen in scanning tunneling microscopy (STM) studies of \bscco.  Charge density modulations were originally observed in the ``halo'' regions around magnetic vortices.\cite{hoff02}  More recent work has provided evidence that these modulations also correspond to PDW order.\cite{edki19}  To appreciate why PDW order might be favored near vortex cores, consider the comparison shown in Fig.~\ref{fg:vc}.  Superconducting order is suppressed within the vortex core, which has a radius equal to the superconducting coherence length, which is of order $4a$, where $a$ is the Cu-Cu lattice spacing.  For an otherwise-uniform superconductor, the superconducting order recovers on the scale of the magnetic penetration depth, which is of order $400a$.  In contrast, the PDW wave function is already modulated with a period of 8a, and has zeros built into it.  If one period is suppressed by a vortex core, it can recover at the edge of the vortex.  This may make PDW order energetically favorable in the region where uniform superconductivity is depressed and slowly recovering.

This case leads to another possible connection.  STM studies\cite{kohs08} have been able to map out the shape of the superconducting gap in reciprocal space from an analysis of quasiparticle interference (QPI) using the ``octet'' model of scattered Bogoliubov quasiparticles.\cite{wang03}  There are two points of interest here. First, to achieve quasiparticle interference one requires defects to cause scattering, such as dilute Zn substitutions for Cu.\cite{pan00a}  Second, the form of the gap obtained from QPI analysis\cite{kohs08} shows differences from that obtained from angle-resolved photoemission spectroscopy.\cite{dama03,vish12}  While the QPI analysis does yield a gap that grows in energy as one moves along the nominal Fermi surface away from the nodal wave vector, it also seems to suggest a gapless arc, centered on the nodal point, that is rather significant for underdoped samples.\cite{kohs08}  This is of interest, because it is exactly the form of gap predicted for a PDW superconductor.\cite{baru08,berg09a}

\begin{figure}[t]
	\centering
	\includegraphics[width=0.9\columnwidth]{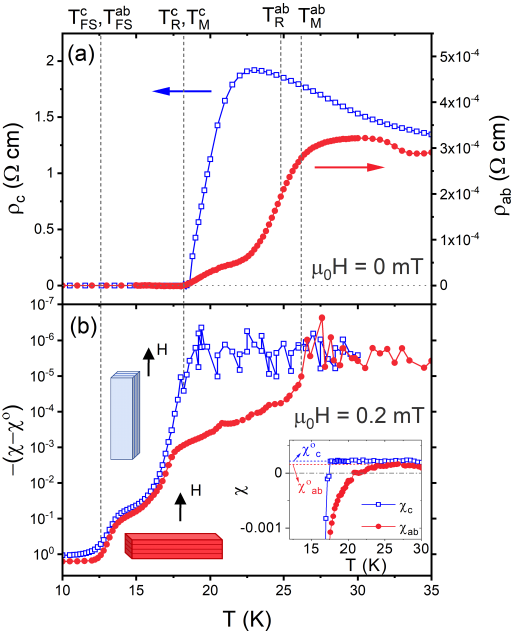}
	\caption{(Color online)  (a) Temperature dependence of the resistivity (in zero field)  for current along the $c$ axis (open squares) and parallel to the $ab$ planes (filled circles). (b) Temperature dependence of the volume magnetic susceptibility with a field of $\mu_0H = 0.2$~mT applied parallel to the planes, $\chi_c$ (open squares), and perpendicular to the planes, $\chi_{ab}$ (filled circles). Inset shows the susceptibilities on a linear scale near the initial transitions; linear fits to the normal state, and extrapolated to low $T$, are indicated by dashed lines. Subtraction of these fits enables the logarithmic-scale plots of the main panel. From Ref.~\citen{loza21}.}
	\label{fg:zn}
\end{figure}

A Zn atom has no $3d$ hole, and hence does not couple to the doped holes when substituted for Cu.  Zn substitution is well known to suppress superconductivity,\cite{fuku96} and $\mu$SR studies have shown that each Zn dopant wipes out superfluid density comparable to a magnetic vortex core.\cite{nach96}  Given the similar features of Zn dopants and magnetic vortices,  one might guess that Zn could locally pin PDW order, so that the QPI detected by STM would selectively probe the PDW gap.  A corollary would be that Zn substitution into superconducting LBCO should lead to 2D superconductivity in zero magnetic field.

To test this possibility, a crystal of LBCO $x=0.095$ with 1\%\ Zn substituted for Cu was studied.  Earlier work had demonstrated that the Zn enhances the spin stripe order, while reducing $T_c$ for 3D order.\cite{wen12a}  Measurements\cite{loza21} of anisotropic resistivity and magnetic susceptibility are shown in Fig.~\ref{fg:zn}.  As one can see, $\rho_{ab}$ shows a substantial drop at $\sim25$~K, together with an onset of weak diamagnetism for magnetic field perpendicular to the planes.  In contrast, $\rho_c$ and the corresponding susceptibility (with field parallel to the planes) only show superconductivity below 18~K, with full bulk shielding achieved below 12~K.  These results are consistent with Zn-dopants inducing filamentary 2D superconductivity at temperatures well above the onset of 3D order.  Thus, a connection between Zn dopants and local PDW order is plausible.

{\newr
\section{Comparison with other cuprates}

As mentioned at the end of Sec.~\ref{SCstripes}, uniform $d$-wave superconductivity requires a spin gap below $T_c$, as is observed for LSCO with $x > 0.13$ \cite{chan08,li18}.  For this range of $x$, RSXS measurements indicate that charge-stripe correlations develop in the normal state, but their scattered intensity decays below $T_c$.\cite{wen19,miao21}

In \ybco, short-range charge order develops in the normal state together with the spin gap and weakens below $T_c$.\cite{huck14,blan14}  Similar short-range charge order is detected in \bscco,\cite{dasi14} where STM studies suggest that it is associated with PDW order pinned around local defects.\cite{du20,chou20}   The defects that occur in \ybco\ tend to be associated with the finite order in the Cu-O chains, as detected by X-ray scattering.\cite{zimm03,ricc13}  Given the important role of charge transfer between the chains and planes,\cite{bozi16} it seems likely that the short-range charge order that develops in the planes serves to screen such defects and involves a PDW character.

We already noted that STM has detected PDW correlations in the halo regions around magnetic vortex cores,\cite{edki19} where the applied magnetic field was 8.25~T.  It is interesting to note that a type of 3D charge order develops in underdoped \ybco\ for fields above 15~T.\cite{wu11,gerb15,chan16a,jang16}  At low temperature and high field, an unusual quantum vortex phase is observed that coexists with quantum oscillations.\cite{yu16,hsu21a}  The possibility that PDW order is associated with this charge order has been discussed.\cite{kacm18,capl21}
}

\section{Conclusion}

As we have seen, the formation of charge stripes in 214 cuprates is an emergent response to doping holes into an AF insulator.  While stripe order competes with 3D superconductivity, it coexists with 2D superconductivity.  In fact, it appears that charge stripes, in the form of 2-leg ladders, can be the source of pairing correlations in cuprates.  There are theoretical analyses that are supportive of this perspective,{\newr\cite{jian21b}} but we will have to wait for studies with advanced numerical approaches to reach a firm conclusion.

\begin{acknowledgment}

This work was supported by the U. S. Department of Energy (DOE), Office of Basic Energy Sciences, Division of Materials Sciences and Engineering, under Contract No.\ DE-SC0012704.  
\end{acknowledgment}

\bibliographystyle{jpsj}
\bibliography{lno,theory}

\end{document}